\documentclass[10pt, conference, letterpaper]{IEEEtran}
\usepackage{balance}
\usepackage[ruled]{algorithm2e}
\usepackage{amssymb}
\usepackage{cite}
\usepackage{color}
\usepackage{multirow}
\usepackage{subfigure}
\usepackage{graphicx,times,amsmath}

\ifCLASSINFOpdf

\else

\fi

\begin{document}
	\title{Wi-Fi Teeter-Totter: Overclocking OFDM for Internet of Things} 
	
	\author{\IEEEauthorblockN{Wei Wang{$^{^\dagger}$}, Shiyue He{$^{^\dagger}$}, Lin Yang{$^{^\ddagger}$}, Qian Zhang{$^{^\ddagger}$}, Tao Jiang{$^{^\dagger}$}}
		\IEEEauthorblockA{{$^{^\dagger}$}School of Electronic Information and Communications, Huazhong University of Science and Technology\\{$^{^\ddagger}$} Department of Computer Science and Engineering\\Hong Kong University of Science and Technology\\
			Email: \{weiwangw, shiyue\_he, taojiang\}@hust.edu.cn, \{lyangab, qianzh\}@hust.edu.cn}}	
\maketitle
	
\begin{abstract} 
The conventional high-speed Wi-Fi has recently become a contender for low-power Internet-of-Things (IoT) communications. OFDM continues its adoption in the new IoT Wi-Fi standard due to its spectrum efficiency that can support the demand of massive IoT connectivity. While the IoT Wi-Fi standard offers many new features to improve power and spectrum efficiency, the basic physical layer (PHY) structure of transceiver design still conforms to its conventional design rationale where access points (AP) and clients employ the same OFDM PHY. In this paper, we argue that current Wi-Fi PHY design does not take full advantage of the inherent asymmetry between AP and IoT. To fill the gap, we propose an asymmetric design where IoT devices transmit uplink packets using the lowest power while pushing all the decoding burdens to the AP side. Such a design utilizes the sufficient power and computational resources at AP to trade for the transmission (TX) power of IoT devices. The core technique enabling this asymmetric design is that the AP takes full power of its high clock rate to boost the decoding ability. We provide an implementation of our design and show that it can reduce the IoT's TX power by boosting the decoding capability at the receivers.
\end{abstract}

\section{Introduction}\label{sec:intro} 


We are entering the post-PC era where an ever-larger variety of smart Internet-of-things (IoT) devices including wearables and smart sensors are increasing the demands for low power communication technologies. Wi-Fi has recently become a contender for this regime due to its compatibility with IP networks and wide deployments of hotspots. Growing numbers of wearables in the market have enabled Wi-Fi connection to the Internet without relying on gateways. Many commercial smart watches including Apple Watch have already equipped with built-in Wi-Fi chipsets. Google's latest Android Wear 2.0 release~\cite{androidwear} allows Android smartwatches to directly connect to Wi-Fi networks.




With the proliferation of sensor-equipped smart objects in streets, homes, and offices, IoT connectivity is envisioned to be massive~\cite{cisco,wang2016less}. Thus, Wi-Fi for IoT devices should be not only low power but also spectrum efficient. As a mature and spectrum efficient multiplexing access technology widely adopted in the latest Wi-Fi standards, OFDM continues its adoption in the new Wi-Fi standard for IoT, i.e., IEEE 802.11ah~\cite{wifihalow}, which reuses OFDM frame format that conforms to IEEE 802.11ac to ensure spectrum efficiency.

\begin{figure}[t]
	\center
	\includegraphics[width=0.49\textwidth]{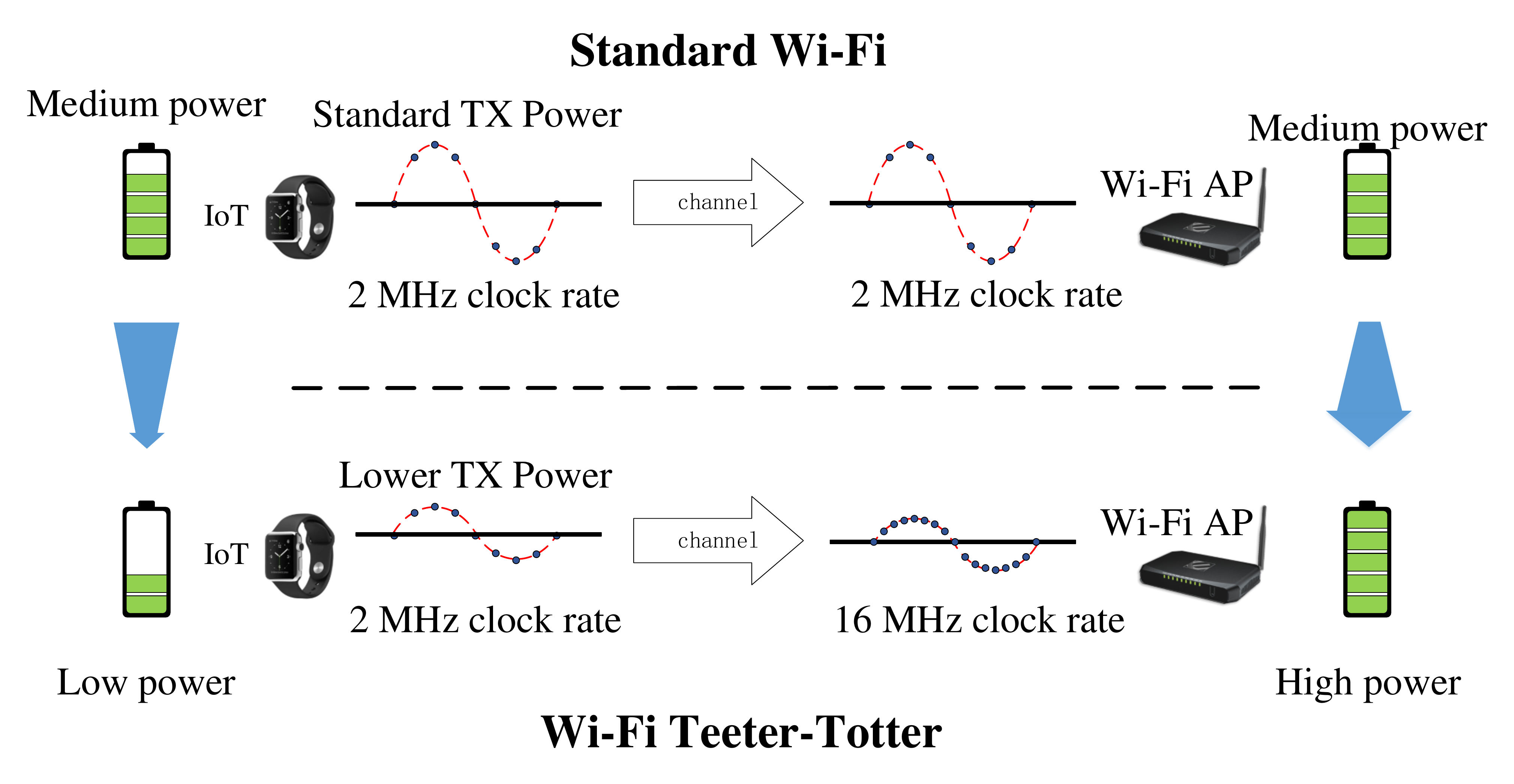}\vspace{0.3cm}
	\caption{The concept of Wi-Fi Teeter-Tooter. A IoT device lowers its TX power and pushes all decoding and energy burdens to the AP side. As the compensation for decreased TX power, the AP takes full advantage of its high clock rate.}\label{fig:rationale}
\end{figure}


A number of recent research efforts have been devoted to reducing the power consumption of OFDM-based Wi-Fi communications. Downclocking receivers' radios is proposed to reduce the power consumption while receiving packets~\cite{wang2017sampleless,mobicom14enfold} or idle listening~\cite{mobicom11emili,wang2017wideband}. Efficient sleep modes~\cite{psm,khorov2015survey} are proposed to reduce the power consumption of conventional IEEE 802.11 nodes by allowing them to enter extremely low power state during idle listening. Despite these growing attempts and extensive efforts, most of them have focused on energy-efficient data reception. A variety of IoT applications require data-rich sensors such as cameras and microphones to frequently upload sampled data to servers, which incurs a large amount of energy consumption for uplink transmission.




In this paper, we argue that the fundamental hurdle for energy efficient uplink transmission lies in the transceiver design. Existing Wi-Fi transceivers are originally designed for symmetric nodes with equal hardware capabilities and power constraints. This assumption no longer stands for IoT applications, where a Wi-Fi AP is much more powerful and almost energy-unconstrained compared to IoT devices. This evokes a similar picture in cellular networks, where base stations are equipped with 100$\times$ sensitive RF front-end and have 20~dB more transmission (TX) power than mobile clients. Analogous to the hardware asymmetry in cellular networks, APs are equipped with Wi-FI chipsets that can support up to 160~MHz bandwidth in IEEE 11ac/11ax, while IoT devices normally support only 1-2MHz bandwidth~\cite{wifihalow}. Thus, there is significant potential to exploit asymmetric PHY configurations to enable low power uplink transmission without undermining decoding performance.

Our line of attack starts from the rationale that we trade the computing and energy resources of APs for the TX power of IoT devices in a \textit{teeter-totter} fashion, that is, we allow IoT devices to transmit uplink packets using the lowest power while pushing all the decoding burdens to the AP side, as illustrated in Figure~\ref{fig:rationale}. In particular, the AP takes full advantage of its clock rate that are tens of times higher than those of IoT's to decode packets with low signal-to-noise ratio (SNR). IoT devices reap benefits from such an asymmetric design by configuring TX power lower than the minimum power required for decoding. On the other hand, the AP with enough computing capabilities and constant power source can afford the extra cost induced by overclocking. Such an asymmetric design trades resources that are cheap to AP for power consumptions that are expensive to IoTs. Such a design completely conforms to Wi-Fi protocols, and thus can be readily integrated with existing standards. The only changes are standalone update of computational logics and RF settings at APs.


A key challenge in realizing the teeter-totter design is how to effectively leverage overclocking in OFDM to decode conventionally undecodable packets under low SNR conditions. Our fundamental insight is that APs yield correlated signals by exploiting the \textit{time shift} effect between overclocked samples. Specifically, when the AP sets its analog-to-digital converter (ADC) clock at a much higher rate than IoT devices, it yields multiple interpolated samples from one transmitted sample. These interpolated samples can be considered as time-shifted versions of the transmitted sample. When transformed into the frequency domain, time-shifted samples result in different phase rotations at different subcarriers. Thus, we can leverage this phase rotation effect to combine these samples for packet decoding. Another challenge stems from the lack of knowledge in modeling the noise distributions between these redundant samples. Instead of blindly reusing conventional decoders, we turn to a data-driven approach. In particular, we build a noise map from preambles without making distribution assumptions. Then, we combine redundant samples by employing a maximum likelihood (ML) decoder based on the joint probability of these samples.

We implement the above teeter-totter design, referred to as \texttt{T-Fi}, on the GNURadio/USRP platform. Evaluation results validate \texttt{T-Fi} in reliably receiving and decoding Wi-Fi packets at low SNR across a wide range of scenarios. Furthermore, \texttt{T-Fi} can reduce the IoT's TX power when the AP is overclocked by a factor of up to 8, which is still lower than the rate used for IEEE 802.11a/g/n.

The contributions of this paper are summarized below.
\begin{itemize}
	\item We provide a thoughtful study towards enabling low-power OFDM transmissions for IoT Wi-Fi standards. Our solution can be seamlessly integrated into existing Wi-Fi standards without modifying the legacy frames or protocols. 
	\item We explore the fundamental structure of overclocked reception in OFDM, and propose a reception pipeline to decode legacy packets at lower SNRs than the conventional transceivers. The key underlying technique is a new decoding algorithm that exploits the time shift effects in oversampled signals.
	\item We build a full prototype of \texttt{T-Fi} and quantify the merits of our design in a wide range of scenarios.
\end{itemize}

The remainder of the paper is structured as follows. We begin in Section~\ref{sec:motivation} with the design motivation. Section~\ref{sec:overclock} analyzes the effect of overclocking in OFDM, which is the underpinnings of our design. Section~\ref{sec:design} elaborates the detailed reception pipeline of our design. System implementation and experimental evaluation are introduced in Section~\ref{sec:implementation} and Section~\ref{sec:evaluation}, respectively. Section~\ref{sec:relatedwork} gives a brief survey of related work, followed by conclusion in Section~\ref{sec:conclusion}.

\vspace{0.3cm}
\section{Motivation of Transceiver Asymmetry}\label{sec:motivation}
In this section, we ask and answer questions to explore practical and suitable transceiver architectures. Through this exploration, we hope to convince readers that the teeter-totter design is a practical design in that it elegantly fits the architecture of IoT Wi-Fi while imposing affordable costs.
\subsection{IoT Scenarios and Requirements}
It is envisioned that the number of smart devices that need to wirelessly share data or access the Internet is growing exponentially. To embrace the coming wave of \textit{massive numbers} of wearables, driverless cars, smart sensors, IEEE 802.11ah was announced in 2016 to tailor Wi-Fi protocols specially for these IoT devices. The target envisioned in this IoT Wi-Fi standard is to challenge Bluetooth and cellular networks by achieving the best of both worlds: to enable \textit{long-distance} communications using relatively \textit{low amounts of power}. Specifically, IEEE 802.11ah defines the following requirements: (i) more than a fourfold number of devices supported by one AP compared to legacy Wi-Fi, (ii) doubled transmission range compared to legacy Wi-Fi links, and (iii) low energy consumption with guaranteed data rates of at least 100~Kb/s.

To meet the above requirements, IEEE 802.11ah introduces a bundle of new features in PHY/MAC. Particularly, the new standard reuses the formerly-adopted OFDM waveform consisting of 32/64 subcarriers while tailoring it to a narrowband channel of 1~MHz or 2~MHz in 900~MHz unlicensed spectrum. OFDM is used to maintain spectrum efficiency for massive connectivity, and narrowband transmission at lower frequencies are used to facilitate low-power, long-range transmissions. In addition, a set of MAC mechanisms is designed to reduce the power consumption in channel access.

\subsection{Hardware Asymmetry}
Although the new standard specifies new PHY and MAC techniques and features to meet the requirements of IoTs, Wi-Fi transceivers still work in a conventional fashion: today's Wi-Fi radios work in a symmetric manner in that the transmitter and receiver are configured with the same clock rate. This symmetric transceiver design has worked well over the past decade as both sides are expected to be (more or less) equally powerful in conventional scenarios. While this assumption no longer stands in IoT scenarios where the disparity between the hardware capabilities of APs and IoTs makes such a design very inefficient. It is either an overwhelming burden for IoTs to match the radio configurations of APs, or makes APs largely under-utilize their hardware capabilities to shoehorn the proper settings of IoTs' radios.

The above situation evokes a similar picture in cellular networks, where base stations are much more powerful than mobile stations and are designed to support thousands of hardware-constrained mobile devices. Cellular networks equip base stations with higher TX power, RF sensitivity, and large form-factor antennas to compensate the hardware gap. The wisdom of cellular networks motivates us to design an asymmetric PHY that fully reap the benefits of the hardware gap between APs and IoTs. 

Our target is to explore an asymmetric transceiver design that can better fit the IoT scenarios and be seamlessly integrated into existing and future OFDM-based Wi-Fi protocols. The fundamental insight is that APs have already equipped with high clock-rate radios to support high-speed Wi-Fi such as IEEE 802.11ac/ax, and the clock rates are tens of times to those of IoTs'. When APs use such overclocked rates to sample IoT's transmissions, a large amount of redundancy between these samples can be used to boost the performance of packet reception, which in turn relaxes TX power requirements imposed on IoTs.

\vspace{0.3cm}
\section{Exploiting Overclocking Opportunities}\label{sec:overclock}

\begin{figure*}[t]
	\center
	\includegraphics[width=7in]{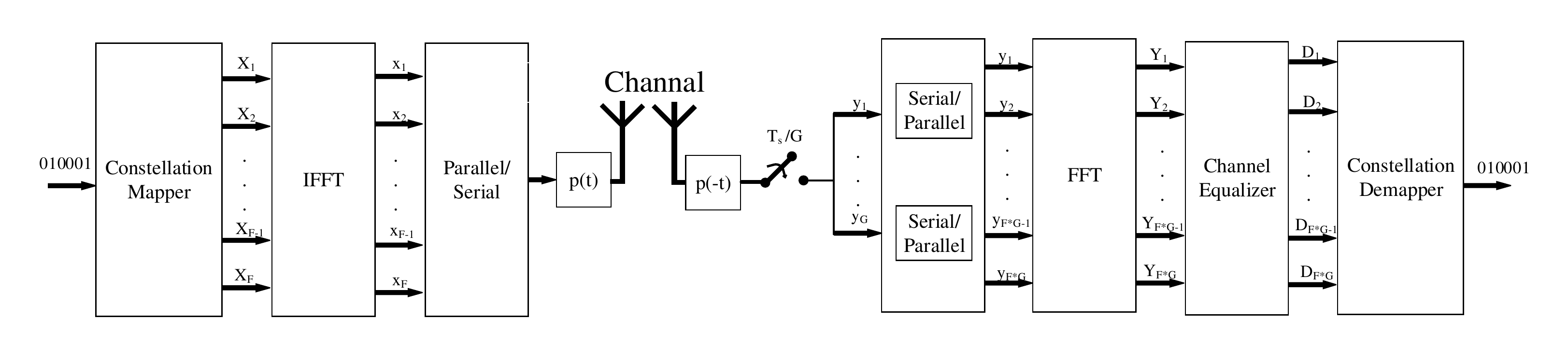}\vspace{0.3cm}
	\caption{OFDM schematic under overclocking.} \label{fig:ofdm_schematic} 
\end{figure*}
At the core of the asymmetric transceiver design, we exploit the redundancy in OFDM signals sampled by overclocked radios. To this end, we discuss the opportunities in retaining the OFDM samples at an overclocked rate and utilizing them to improve symbol decoding. We start by analyzing the correlations among these samples.

In a standard OFDM system, as illustrated in Figure~\ref{fig:ofdm_schematic}, one OFDM symbol contains a sequence of bits that are modulated into a set of lattice points $X[f]$ on orthogonal subcarriers in the frequency domain. The information-bearing $X[f]$. For OFDM transmissions, inverse fast Fourier transform (IFFT) are used to transform $X[f]$ into a time-domain sequence
\begin{equation}
	x[n] = {1 \over F} \sum_{f=0}^{F-1} X[f] e^{j 2\pi f n /F}, n=0,...,F-1,\notag
\end{equation}
where $F$ is the number of subcarriers. Note that we omit cyclic prefix for simple illustration. The discrete-time signal $x[n]$ passes an digital-to-analog converter (DAC) and a pulse shaping filter to transform itself into a continuous-time signal in baseband form, which is written as
\begin{equation}
	x(t) = \sum_{f=0}^{F-1} x[n] p(t-f T_s),\notag
\end{equation}
where $p(t)$ is the impulse response of the pulse-shaping filter and $1/T_s$ is the baud rate. $x(t)$ is transmitted and propagates a wireless channel with impulse response $h(t)$. A receiver receives the signals with a matched filter $p(-t)$, and obtain
\begin{equation}
	y(t) = \sum_{n=0}^{F-1} x[n] p(t-f T_s) \ast h(t-f T_s) \ast p(f T_s-t) + w(t),\notag
\end{equation}
where $\ast$ denotes convolution, and $h(t)$, $w(t)$ are the impulse response of the channel and complex Gaussian noise, respectively. In the presence of a time-dispersive channel and additive noise, the received continuous time-domain baseband signal $y(t)$ can be expressed as
\begin{equation}
	y(t) = {1 \over F} \sum_{f=0}^{F-1} X[f] H(f) e^{j {2\pi f t \over T}  } + w(t),\notag
\end{equation}
where $H(f)$ is the channel frequency response at subcarrier $f$, and $T$ is the symbol duration. 

Now we consider a concrete example in which $F=64$ subcarriers are used to convey information. When using the same clock rate to sample received signal $y(t)$, the sampling instances are at $t={n \over 64} T$. Thus, the receiver yields
\begin{equation}
	y_1[n] = {1 \over 64} \sum_{f=0}^{63} X[f] H(f) e^{j {2\pi f n \over 64}  } + w_1(n).\notag
\end{equation}
When the receiver doubles the clock rate, the number of samples in each FFT segment is 128. As such, the sampling instances are at $t={n \over 128} T$, and the time domain sample sequence is expressed as
\begin{align}
	y[n] =& {1 \over 64} \sum_{f=0}^{63} X[f] H(f) e^{j {2\pi f n \over 128}  } + w(n), n = 0,...,127 \notag\\
	= & {1 \over 64} \sum_{f=0}^{63} \left( X[f] H(f) e^{j {2\pi f  \over 128} 2n } + X[f] H(f) e^{j {2\pi f  \over 128} (2n+1) }  \right) \notag\\
	&+ w_1(2n) + w_1(2n+1), n = 0,...,63 \notag\\
	= & y_1[n] + y_2[n] + w_2(n), n = 0,...,63, \notag\\
\end{align}
where $y_2[n] = {1 \over 64} \sum_{f=0}^{63} X[f] H(f) e^{j {2\pi f  \over 64} (n+1/2)}$ and $w_2(n)$ is the new noise sequence in oversampled signals. We see that the oversampled signal $y[n]$ consists of two polyphase components $y_1[n]$ and $y_2[n]$. Notice that $y_2[n]$ is a time-shifted version (delayed by a half sample) of $y_1[n]$. The frequency response of $y_1[n]$ can be derived by performing FFT.
\begin{align}
	Y_1[l] &= \sum_{n=0}^{63} y_1[n] e^{-{j 2\pi n l \over 64}}, l=0,...,63 \notag \\
	&= {1 \over 64} \sum_{n=0}^{63} \sum_{f=0}^{63} X[f] H(f) e^{j {2\pi f n \over 64}  } e^{-{j 2\pi n l \over 64}} +  W_1(l) \notag \\
	&= X[l] H(l) + W_1(l),\notag
\end{align}
where $ W_1(l)$ is the noise expressed in the frequency domain. Similarly, the frequency response of $y_2[n]$ is expressed as
\begin{align}
	Y_2[l] &= \sum_{n=0}^{63} y_2[n] e^{-{j 2\pi n l \over 64}}, l=0,...,63 \notag \\
	&= {1 \over 64} \sum_{n=0}^{63} \sum_{f=0}^{63} X[f] H(f) e^{j {2\pi f  \over 64} (n+1/2)} e^{-{j 2\pi n l \over 64}} +  W_2(l) \notag \\
	&= X[l] H(l) e^{-{j \pi l \over 64}} +  W_2(l).\notag
\end{align}
Ignoring the noises, $Y_2[l]$ is a phase-shifted version of $Y_1[l]$, and the amount of phase rotation is linear to the subcarrier frequency $l$. 

\begin{figure}[t]
	\center
	\includegraphics[width=3.5in]{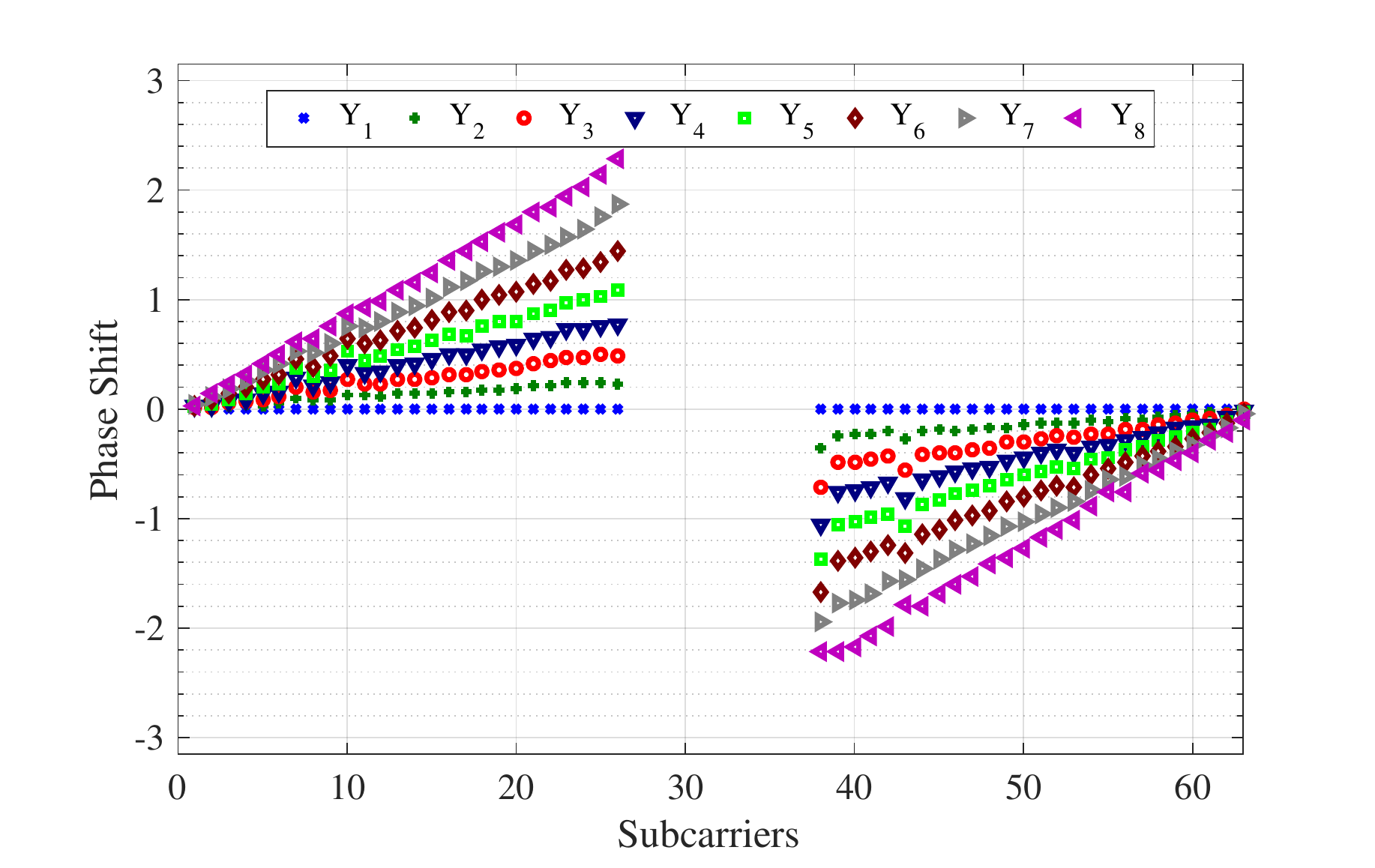}\vspace{0.3cm}
	\caption{Phase shift induced by oversampling a real Wi-Fi packet. Phases are normalized according to the phases of $Y_1$. Subcarriers 0-25, 38-63 carrying payload data are compared.} \label{fig:phaseshift} 
\end{figure}

Hence, doubling the receiver's clock rate yields two receptions, which transforms the system into a single-input-multiple-output (SIMO) system. When we employ $G$-fold clock rate at the receiver, we can obtain $G$ phase-shifted versions of $Y_1[l]$ described as
\begin{equation}\label{e:yg}
	Y_g[l]	= X[l] H(l) e^{-{j 2 \pi gl \over 64G}} +  W_g(l), g=0,...,G-1.
\end{equation}
Figure~\ref{fig:phaseshift} shows the phase shifts across all subcarriers in a real Wi-Fi packet when received at eight-fold clock rate using USRP testbeds. The predictable phase shifts in $Y_g[l]$ can be easily compensated to obtain $G$ copies of the transmitted signal. We can combine these copies to improve the received signal strength while amortizing noise, and thereby enhance the decoding capability under low SNR conditions.

\vspace{0.3cm}
\section{Asymmetric Transceiver Design}\label{sec:design}
In this section, we describe our design of \texttt{T-Fi}, an asymmetric transceiver architecture for IoT Wi-Fi. \texttt{T-Fi} can fully interoperate with standard IEEE 802.11 devices, with no modifications to existing protocols. \texttt{T-Fi} leverages the high clock rate of APs to enable low power transmissions for IoT devices. This section elaborates the detailed reception pipeline design that embraces this design rationale.
\subsection{Frame structure and reception pipeline}
A Wi-Fi frame starts with a preamble prepending payload to perform time synchronization, carrier frequency offset (CFO) compensation, and channel estimation. In particular, a preamble consists of a Short Training field (STF) and a Long Training Field (LTF). The legacy STF contains two OFMD symbols that are comprised of ten repetitions of a 16-sample sequence, and the LTF contains two identical 64-sample (80-sample including cyclic prefix) OFDM symbols. A PHY header, known as signaling field (SIG), containing modulation and coding scheme (MCS) information, duration, and service field, sits between the preamble and payload. The preamble and the first 24 bits of the PHY header are encoded using the lowest MCS mode to ensure their correct reception, while the rest can be encoded using different MCS modes based on a certain rate adaptation algorithm.

Under overclocking settings, a receiver increases the sampling rate of the ADC by $G$ folds. The incoming analog signals are oversampled and converted into $G$ baseband streams, which are fed into the synchronization block to identify the start of a Wi-Fi frame. Meanwhile, the CFO between the sender and the receiver is estimated and compensated using preamble sequences to eliminate the effect of the disparity between clock oscillator frequencies. Since there may still be residual frequency offset, phase compensation is employed to correct the phase rotation throughout the entire payload. In Wi-Fi, four subcarriers are used as pilots to estimate phase rotation in each data symbol. Then, the receiver performs FFT on the samples to obtain their frequency domain values. After performing FFT, the receiver estimates channel responses by comparing the received values in LTF and the actual values transmitted. Finally, the receiver decodes the bits in the payload symbols. 

In what follows, we systematically exploit the merits as well as issues brought by overclocking, and propose corresponding techniques to optimize the packet reception performance.

\subsection{Timing Synchronization}
Timing synchronization aims to detect the presence of a Wi-Fi packet and then identify the start of Wi-Fi symbols. Wi-Fi receivers take advantage of the periodicity in STF to detect the Wi-Fi preamble. It measures the energy of the sampled signal $\mathbf{y}$: $P = \sum_{k=1}^L \left|y[k]\right|^2$, where $L$ is the measuring window. If $P_s$ is higher than a given threshold, it then performs auto correlation to check whether the incoming signal is a Wi-Fi frame. Specifically, it tracks the auto correlation of the signal with $d$ delayed samples:
\begin{equation}
	c_{\text{auto}}=\sum_{k=1}^{L} y[n+k] r^{*}[n+k-d].\notag
\end{equation}
$d$ is normally set to be 16, which is the period of one short sequence in STF. For a Wi-Fi packet, $c_{\text{auto}}$ spikes and forms a plateau. Our experiments show that even at very low SNR, the plateau is still clear to be separated from noise. Therefore, we configure normal clock rate for packet detection to avoid unnecessary energy consumption in idle listening and carrier sensing.

\begin{figure}[t]
	\center
	\includegraphics[width=3.5in]{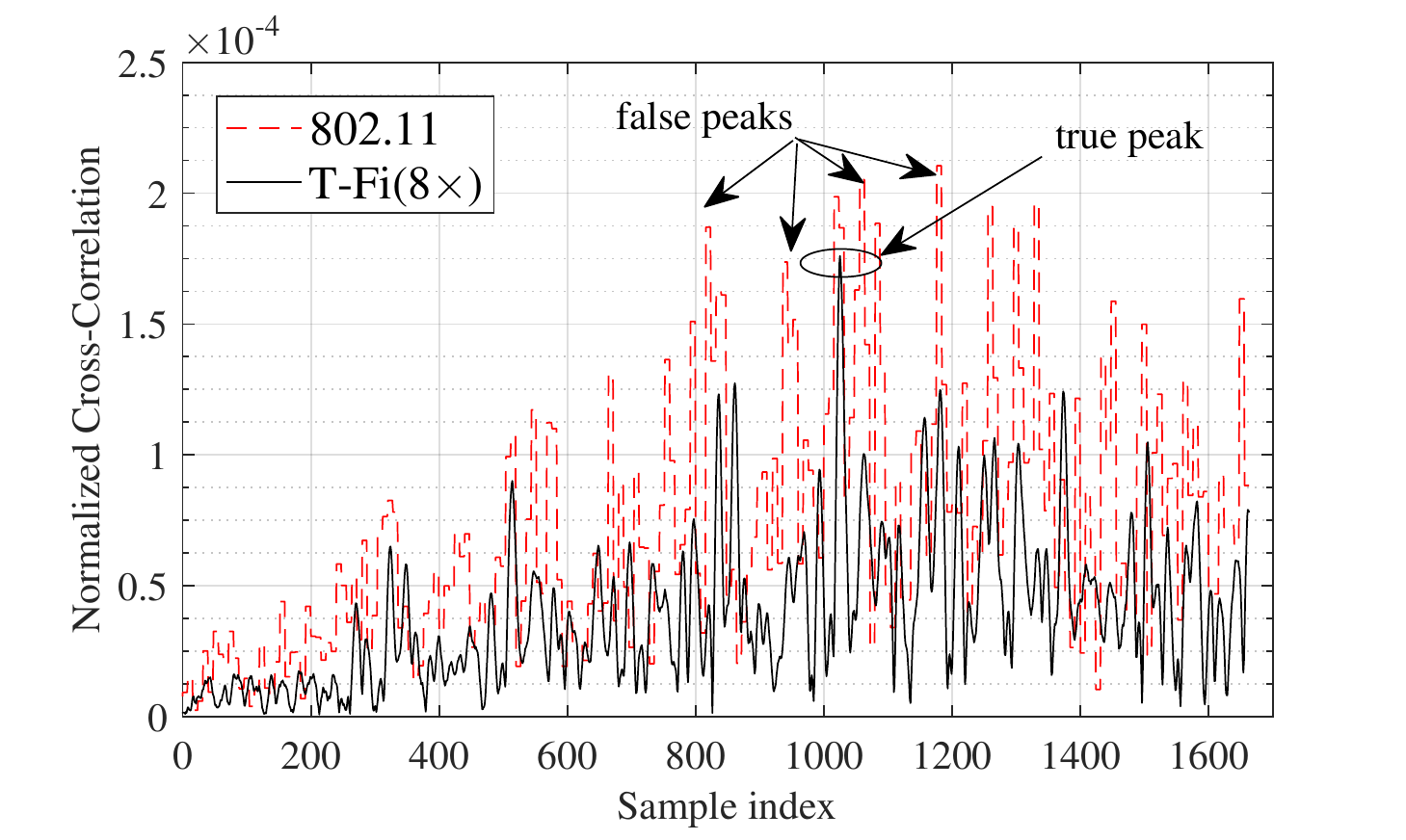}
	\caption{Cross correlation responses at various clock rates. We set up a USRP link to capture raw data samples pf a real Wi-Fi packet at 8$\times$ and 1$\times$ Nyquist rates, respectively.}\label{fig:sync_eg} 
\end{figure}

After successfully detecting a Wi-Fi packet, the receiver precisely locates the boundary between OFDM symbols. While the IEEE 802.11 specifications do not mandate any specific algorithm, a typical synchronization algorithm is to use the cross-correlation property of the LTF. In particular, the receiver correlates the received signal with the transmitted time-domain LTF sequence. However, when the SNR is poor, cross correlation cannot cancel out noise, thereby resulting in multiple false peaks. We exploit correlations in oversampled LTF signals to overcome this predicament. The standard LTF consists of 52 subcarriers with the identical magnitude and $0$ or $\pi$ phase. When the LTF is sampled by an overclocked ADC, the correlation between oversampled data can be described by Eq.~\ref{e:yg}. Hence, the receiver can treat all the samples as a new LTF, which has stronger correlation properties. Figure~\ref{fig:sync_eg} illustrates the merits of using the oversampled LTF for synchronization. Under low SNR conditions, the cross correlation result of a standard LTF produces multiple comparable peaks, which makes the receiver easily aligns to a wrong peak. The cross correlation result of an oversampled LTF produces a single highest peak corresponding to the full alignment of OFDM symbols.

\begin{figure}[t]
	\center
	\includegraphics[width=3.5in]{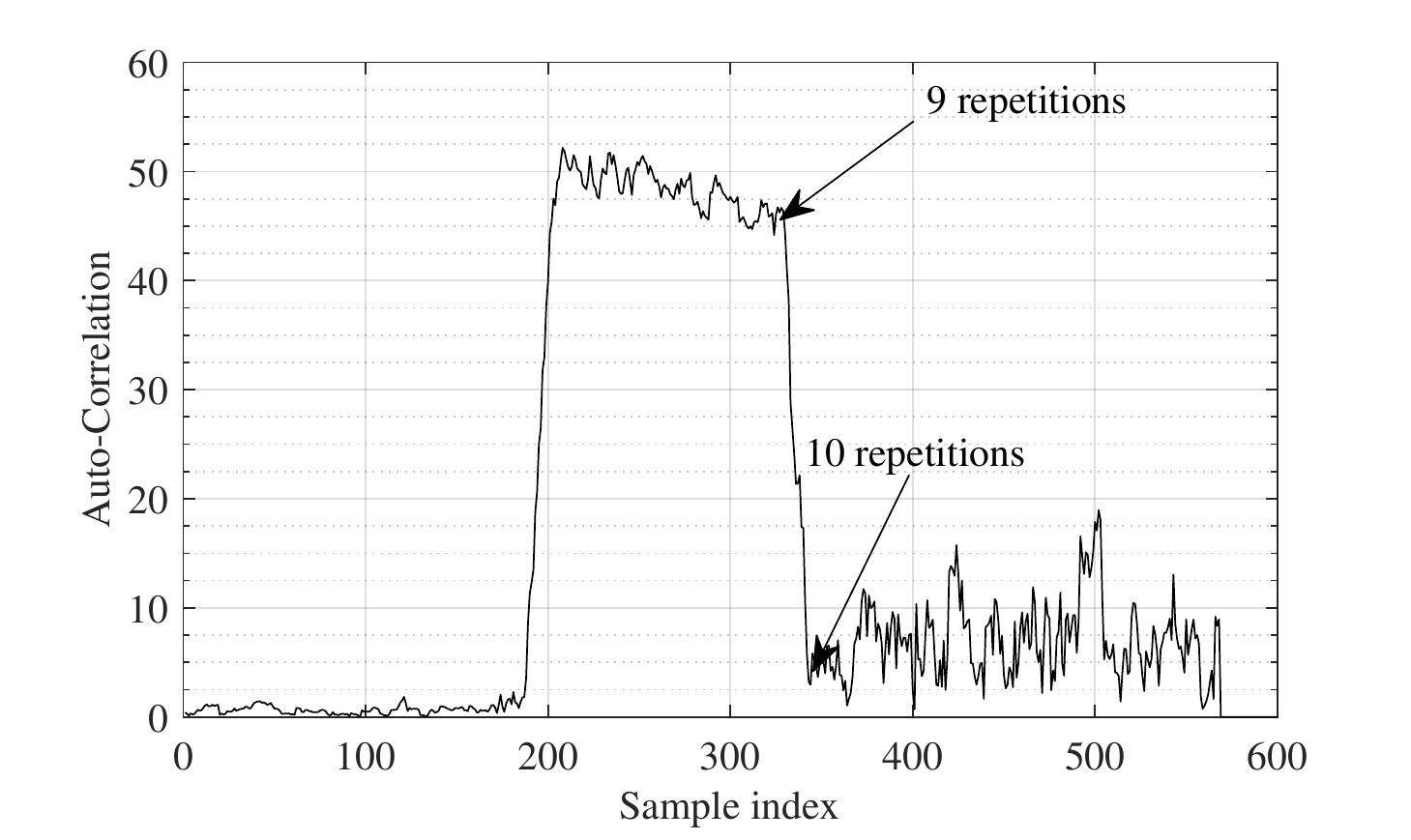}\vspace{0.3cm}
	\caption{Auto correlation response.} \label{fig:autocorr} 
\end{figure}

We set normal clock rate for idle listening to avoid unnecessary energy consumption. After packet detection, the receiver switches to the overclocking mode, which incurs latency for clock switch. As the receiver keeps the same frequency synthesizer and the center frequency of its analog circuit, the latency comes from the digital phase-locked loop (PLL) stabilization. Wi-Fi radios take merely several microseconds (e.g., 8~$\mu s$ in MAXIM2831~\cite{maxim}) to stabilize PLL. As the state-of-the-art IoT Wi-Fi, i.e., IEEE 802.11ah, shrinks the conventional 20~MHz bandwidth to 2~MHz while retaining the same number of subcarriers, the duration of one OFDM symbol is extended to 40~$\mu s$. Both STF and LTF last 80~$\mu s$, which leaves enough time for clock switch. As illustrated in Figure~\ref{fig:autocorr}, which is tested with SNR=10 dB, we observe that the first nine repetitions are enough to produce a plateau for packet detection. Thus, we take the first nine repetitions in STF to perform auto correlation while leaving the last repetition of STF for clock switch.In our experiment results, we show that it does not need to extra packet loss and the whole timing synchronization perform is boosted compared to the standard algorithm under low SNR conditions.


\subsection{Frequency Offset Compensation}
CFO varies over time and must be estimated and compensated for each frame. In practice, the received baseband signal, instead of being centered at DC (0 Hz), is centered at a frequency offset $\Delta f$.
\begin{align}
	y_{\text{CFO}}(t) &= y(t)e^{\frac{j2\pi\Delta ft}{F_{s}}},\notag
\end{align}
where $F_s$ is the sampling frequency. CFO induces phase rotation over time that not only undermines the payload decoding but also affects the phase correlation among $Y_g[l]$. As our overclocking design relies on the phase correlation, CFO must be precisely estimated and calibrated.

Standard Wi-Fi receiver estimates CFO by comparing the phase rotation between the two identical OFDM symbols in LTF. Such an estimation is accurate enough for normal packet decoding, while the residual offset cannot be neglected when using the phase correlation among oversampled data. We make a fine-grained calibration in the following data symbols by exploiting the CFO effect among oversampled sequences $y_g[n]$. Let $\mathbf{y}_g = [y_g[0],...,y_g[N-1]]^\top$ denotes $g$th copy of the oversample signals, and the received data values in frequency domain $\mathbf{d} = [X[0]H(0),...,X[N-1]H(N-1)]^\top$. FFT is expressed in the form of a $N \times N$ matrix $\mathbf{F}$, where each entry $f_{ij}=e^{j (i-1)(j-1)}$. The effect of CFO on the first copy can be expressed as $\mathbf{P}=\text{diag}\left(1 \; e^{j {\Delta f\over F_{s}}} \cdot \cdot \cdot e^{j {\Delta f (N-1)\over F_{s}}}\right)$. The phase shifts due to overclocking can be described by $\mathbf{O}_g=\text{diag}\left(1 \; e^{-{j 2 \pi g \over NG}} \cdot \cdot \cdot  e^{-{j 2 \pi g(N-1) \over NG}} \right)$. Then, the $g$th copy of the oversample signals $\mathbf{y}_g$ can be expressed as
\begin{equation}
	\mathbf{y}_g = e^{\frac{j2\pi\Delta fg}{F_{s}G}} \mathbf{P} \mathbf{F} \mathbf{O}_g \mathbf{d} + \mathbf{W_g}.
\end{equation}
In the absence of noise, we obtain the following the relationship
\begin{align}\label{e:cfo}
	\mathbf{F}^{\text{H}} \mathbf{P}^{\text{H}} \mathbf{y}_g &= e^{-\frac{j2\pi\Delta fg}{F_{s}G}}  \mathbf{O}_g^{\text{H}} \mathbf{F}^{\text{H}} \mathbf{P}^{\text{H}} \mathbf{y_0},
\end{align}
where $(\cdot)^{\text{H}}$ is the conjugate transpose of a matrix. It is intuitive to find the unique $\Delta f$ that satisfying the above equation. In the presence of noises, we minimize the sum of distances between the left and the right hands of Eq.~\ref{e:cfo} for all $g>0$. We employ ML estimator to derive $\Delta f$.

\subsection{Decoding}
After timing synchronization and frequency offset compensation, channel response needs to be understood before packet decoding. Conventional Wi-Fi receivers utilize the two training symbols in LTF to estimate the channel response at each subcarrier. As the data bits encoded on each subcarrier in LTF are pre-known by the receiver, the channel response is obtained by comparing the received data values and the transmitted ones. Based on the estimated channel response, the receiver removes the effect of wireless channel on the following OFDM symbols to extract the transmitted data values. Noises are assumed to be independent and Gaussian distributed, and thus the receiver demodulates each data value by mapping it to the nearest lattice point in the constellation map. We extend the above rationality to the overclocking case by averaging multiple copies of compensated samples before feeding them into the standard ML decoder.

\subsection{Putting Everything Together}
\begin{figure}[t]
	\center
	\includegraphics[width=3.5in]{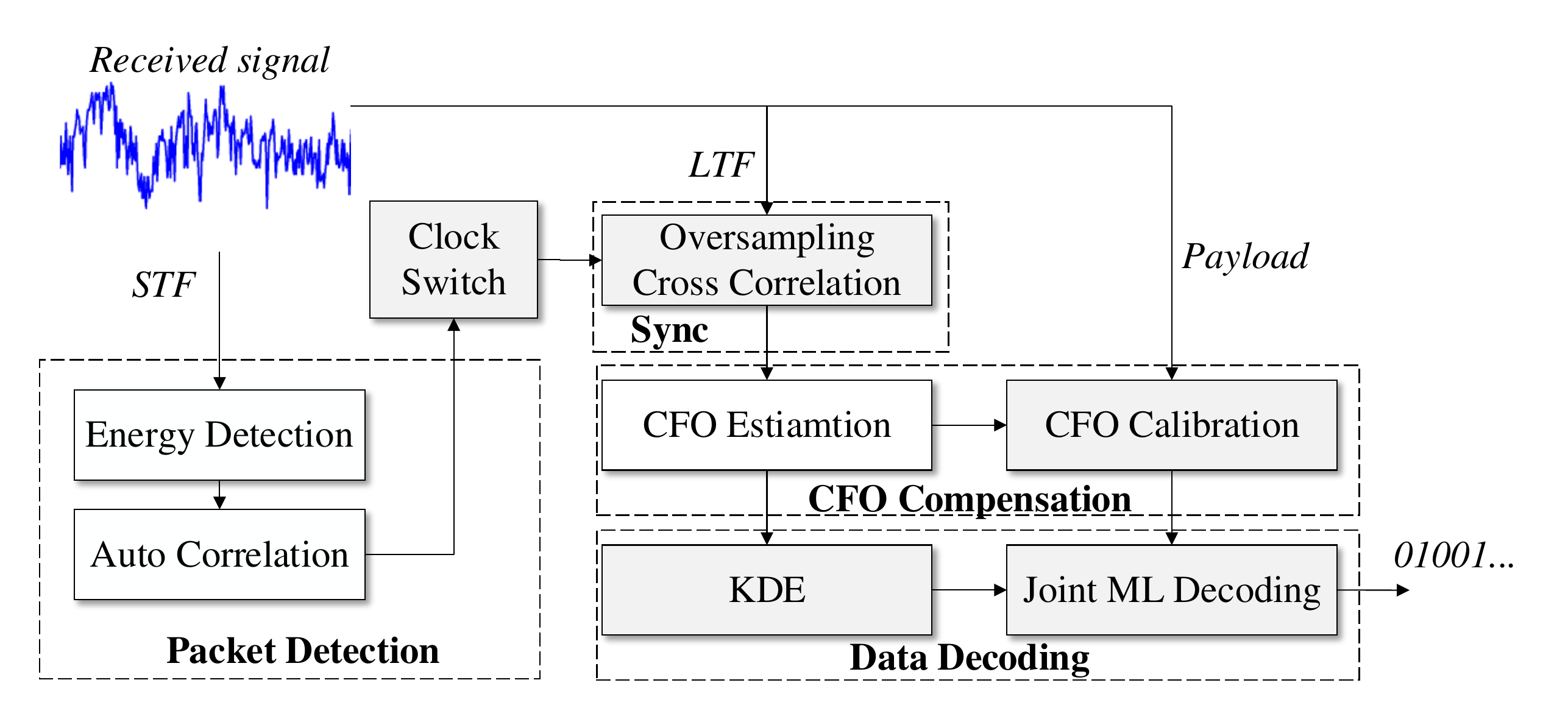}\vspace{0.3cm}
	\caption{System architecture of \texttt{T-Fi}.} \label{fig:architecture}\vspace{0.3cm}
\end{figure}

An overview of \texttt{T-Fi} system architecture is shown in Figure~\ref{fig:architecture}. \texttt{T-Fi} leverages the AP's high clock rate and processing capability to grant lower TX power for IoT devices. The reception pipeline conforms to the logic of a standard Wi-Fi receiver but is tailored to the asymmetric design to make the most of the redundancy induced by overclocking. An AP uses the first nine repetitions of the STF for packet detection and then switches to the high clock rate mode within the STF reception within the last one repetition. As the symbol length in IoT Wi-Fi packets is 10$\times$ of that in conventional Wi-Fi packets, it leaves enough time for the state-of-the-art chips to switch and stabilize the clock. Under the overclocking setting, the AP yields redundant LTF and payload samples, and exploits the correlations among these samples to boost the performance of synchronization and CFO estimation under low SNR conditions. Then synchronization algorithm is performed on the LTF, while CFO estimation is performed on the LTF and then calibrated in the following data symbols in the payload. Multiple copies of oversampled samples are averaged and fed to the standard ML decoder.


\section{Implementation}\label{sec:implementation}
\texttt{T-Fi} can be realized in the existing OFDM PHY with no hardware changes. We prototype \texttt{T-Fi} on top of the OFDM implementation on the GNURadio/USRP platform. We implement the entire PHY design specified in Section~\ref{sec:design} directly in the USRP Hardware Drive (UHD). We use USRP B210 nodes connecting to PCs with Intel i7 quad-core processor and 8~GB memory for the testbed setup. Nodes in our experiments are configured to operate in the 2.4-2.5~GHz or 900~MHz range. 

The Transmitter is configured to continuously send Wi-Fi packets conforming to IEEE 802.11ah PHY format. It operates on a 2~MHz or 1~MHz channel, of which 52 subcarriers are configured to carry data values while 4 subcarriers are pilot tones. The symbol duration is 40~$\mu s$. We adopt the legacy PHY layer convergence procedure (PLCP) format of IEEE 802.11ah, where the PLCP preamble consists of 2 STF OFDM symbols and 2 LTF OFDM symbols. There are 10 MCSs starting from 1/2 BPSK to 5/6 256QAM. Due to hardware limitations, we only implement BPSK, QPSK, 16QAM, and 64QAM modulations.


\begin{figure}[t]
	\center
	\includegraphics[width=3.5in]{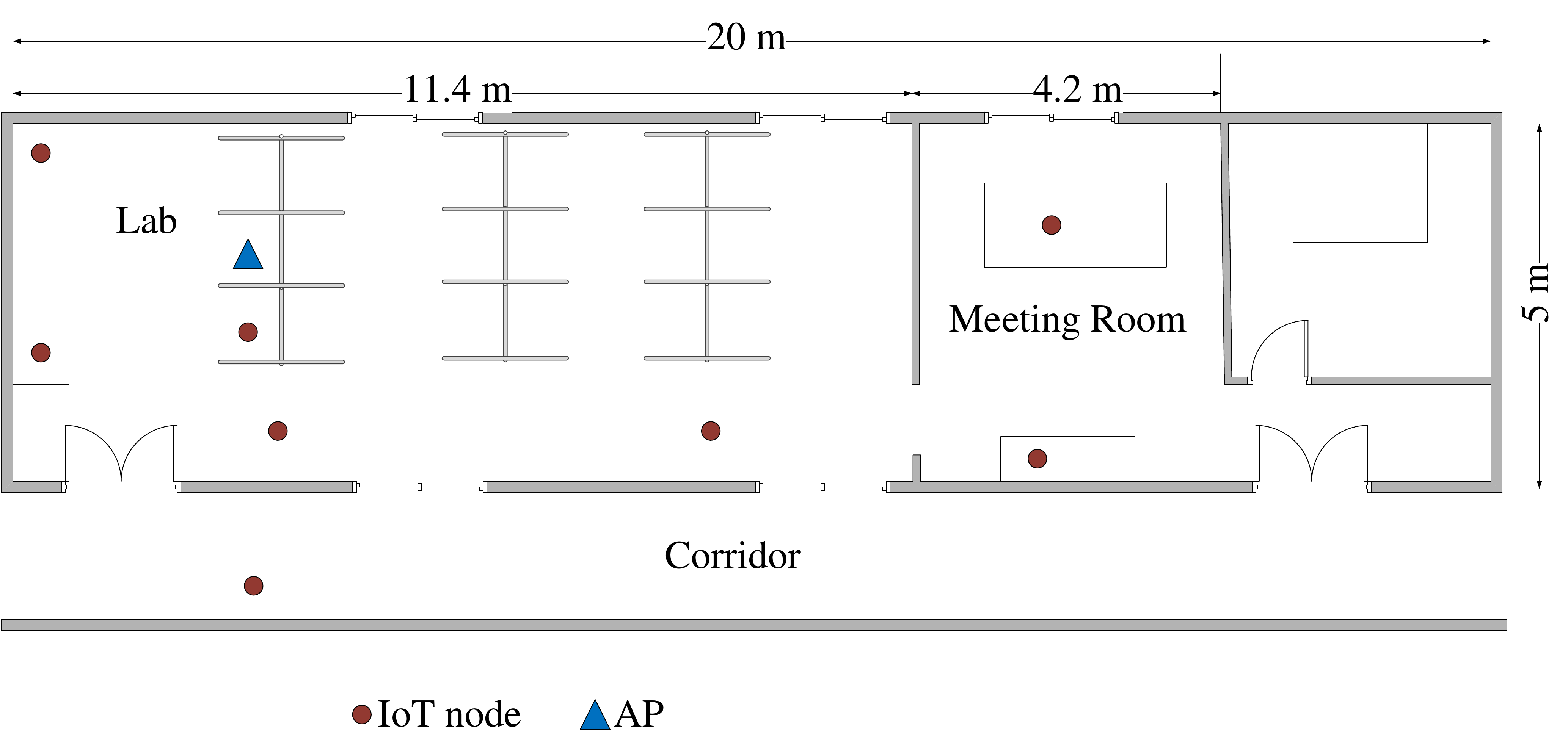}\vspace{0.3cm}
	\caption{Experimental floorplan.} \label{fig:floorplan} 
\end{figure}
\section{Evaluation}\label{sec:evaluation}
In this section, we present a detailed experimental evaluation of \texttt{T-Fi}. Our experiments center around two questions: (i) How much decoding performance improvement can \texttt{T-Fi} provide in real wireless environments under proper overclocking settings? (ii) How much TX power can \texttt{T-Fi} save without compromising the decoding performance? To answer these questions, we conduct a set of experiments to evaluate the performance of synchronization, decoding, the overall packet reception, as well as TX power.


\subsection{Experimental Setup}

We conduct our experiments using USRP B210 nodes deployed in an indoor environment with metal/wood shelves and stone walls. Figure~\ref{fig:floorplan} illustrates the floorplan of the environment, where nodes are placed in a lab, a meeting room, a hallway, and a corridor. We control the TX power by adjusting the programmable-gain amplifier (PGA) in USRP. As such, we sweep the SNR range from 9~dB to 35~dB. For each setting, a USRP node sends approximate 3000 packets. For each setup, we vary the clock rates of the receiver from the normal clock rate ($1\times$) to overclocked rates ($2\times$, $4\times$, $8\times$). Note that the clock rate of IEEE 802.11ac/ax receiver is $10$-$80\times$. We use the standard IEEE 802.11 packet reception at the normal clock rate as the baseline for comparison.

%
\subsection{Synchronization}

\begin{figure}[t]
	\center
	\includegraphics[width=3.5in]{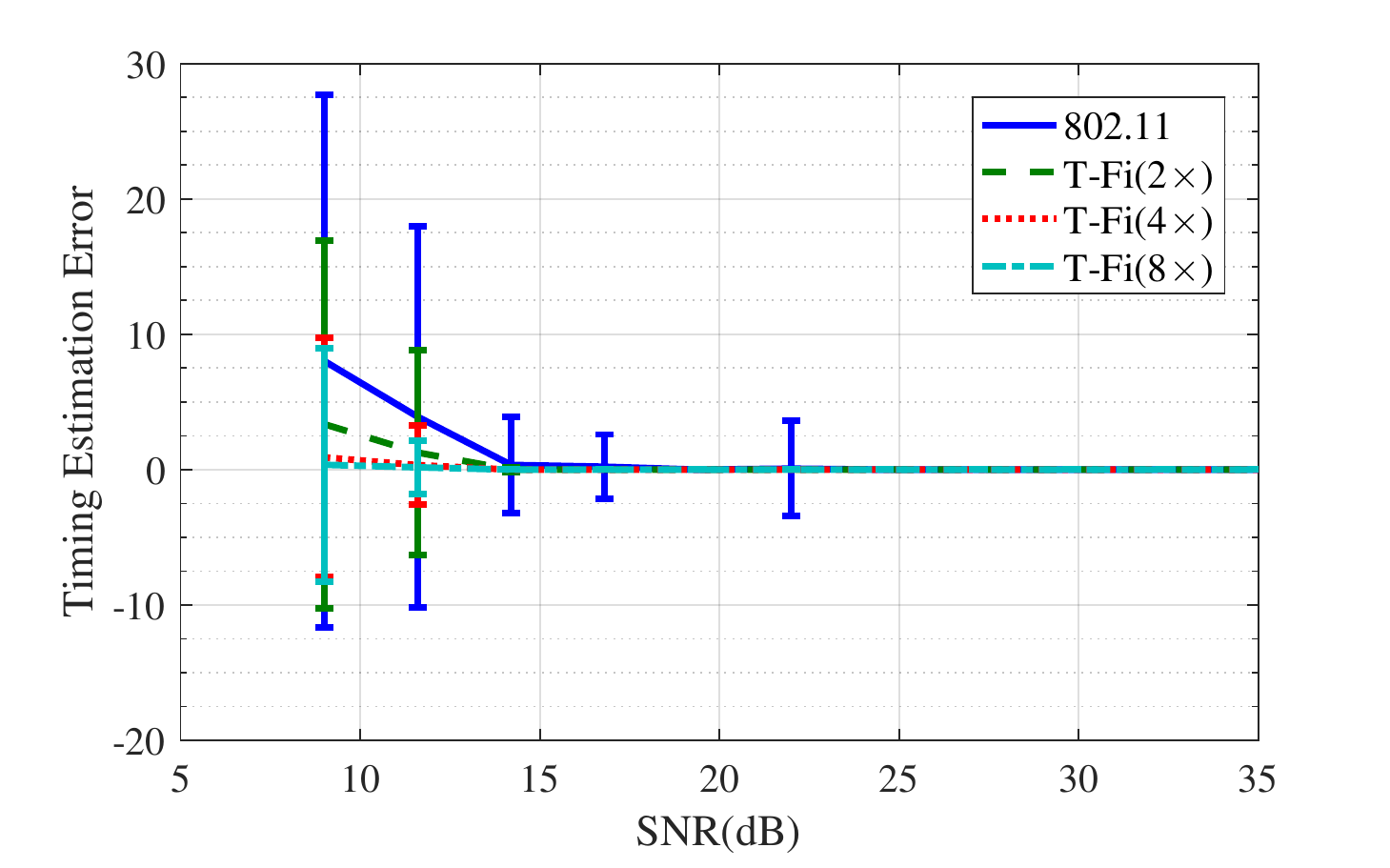}\vspace{0.3cm}
	\caption{Synchronization error.} \label{fig:sync} 
\end{figure}
First, we show that \texttt{T-Fi} addresses the synchronization issue under low SNR conditions. Recall that under low SNR conditions, conventional cross correlation algorithm cannot cancel out noise and yields multiple false peaks. We instead exploit the inherent correlation in the oversampled LFT signals to enhance the correlation property. Since we cannot directly obtain the ground truth of synchronization, we capture raw USRP samples and conduct offline analysis. In particular, we log raw samples received by a USRP node and feed them into an emulator to rehearse the synchronization process. We run the synchronization \texttt{T-Fi} at multiple clock rates (2$\times$, 4$\times$, and 8$\times$ clocking rates) and the standard synchronization algorithm at the normal clock rate (1$\times$ clock rate). 

Figure~\ref{fig:sync} compares the timing estimation error, which is measured in number of samples. Error bars show the standard deviation of time synchronization offset. Under all SNR conditions, synchronization error diminishes with the increment of clock rates. The standard synchronization algorithm works well for SNR$>15$, while the synchronization error grows substantially when SNR is low (9~dB - 12~dB). Empowered by overclocking, \texttt{T-Fi} largely reduces the synchronization error. When SNR=9~dB, the average synchronization error at $8\times$ clock rate is 0.91, which is merely 13\% of the average error at the standard clock rate. These results confirm the merit of using over-clocking to do time synchronization.

\subsection{Decoding}

\textbf{Decoding performance under various SNR conditions.} Our second set of experiments evaluates the decoding performance at different MCS modes under various environments and SNR conditions. To focus on the decoding performance, we use the same synchronization algorithm at the same clock rate for all decoders. In particular, we set $8\times$ clock rate to receive packets, and perform synchronization and CFO compensation. Unless otherwise stated, USRP nodes operate at 900~MHz, which is the frequency band specified in IEEE 802.11ah.



\begin{figure}[t]
	\center
	\includegraphics[width=3.5in]{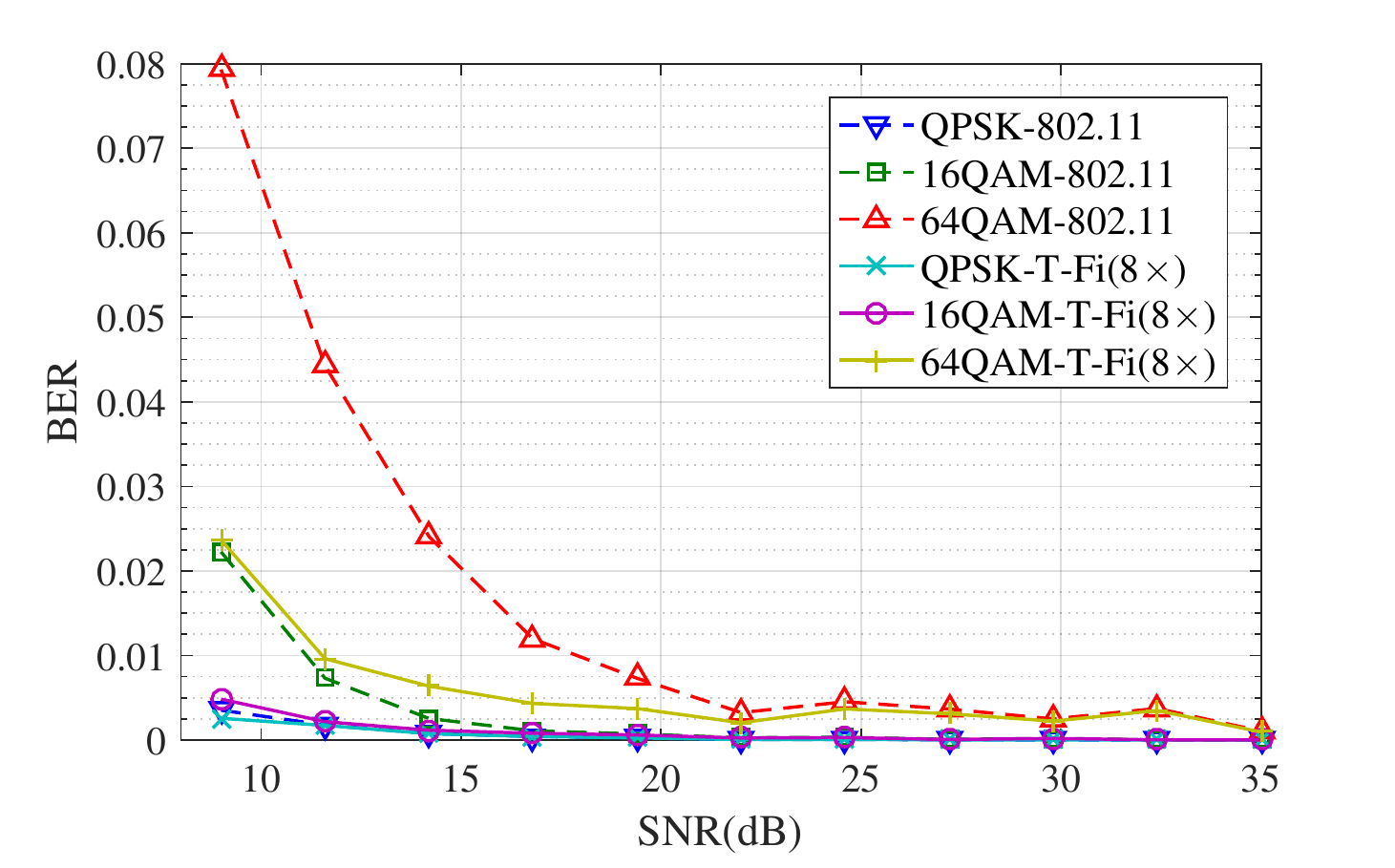}
	\caption{BER for different modulations.} \label{fig:modulation}\vspace{0.1cm}
\end{figure}

In Figure~\ref{fig:modulation}, we compare the decoding performance under different modulation schemes. The clock rate of \texttt{T-Fi} receiver is $8\times$ of the standard 802.11 receiver. As expected, for all modulations demonstrated, \texttt{T-Fi} outperforms the standard 802.11 receiver. We also observe that \texttt{T-Fi} substantially reduces BER compared to the standard receiver for high-order modulations. When SNR=9~dB, the BER of \texttt{T-Fi} is merely 25\% of the standard receiver's BER. The reason is that high-order modulations are more error-prone under low SNR conditions, which leaves more room for overclocking to correct these errors.

Another observation is that the decoding performance of \texttt{T-Fi} for 64QAM (16QAM) is comparable to that of the standard receiver for 16QAM (QPSK). This result implies that \texttt{T-Fi} can use higher-order modulations for packet transmission compared to standard Wi-Fi, and thus delivers roughly $4\times$ data rate. Hence, we can envision a new dimension of the \texttt{T-Fi} is to improve the throughput in IoT Wi-Fi.

\textbf{Impact of wireless environment.} As analyzed in Section~\ref{sec:overclock}, the decoding performance depends on channel response and noise. In our previous experiments, all nodes are in the same lab, and we focus on the decoding performance under different SNR conditions. In this experiment, we evaluate the impact of wireless environments, which lead to different multi-path fading and shadowing. We repeat the previous set of experiments at three additional locations in the same building but with significantly different propagation environments. The locations of IoT nodes and the AP include a hallway in the lab, a corridor outside the lab, and a meeting room next to the lab, as illustrated in Figure~\ref{fig:floorplan}. We use a fixed PGA to set the same TX power for all locations.

\begin{figure*}
	\centering
	\begin{minipage}[b]{0.66\textwidth}\centering
		\subfigure[\scriptsize Different sample rates]
		{\label{fig:loc_rate}\includegraphics[width=0.5\textwidth]{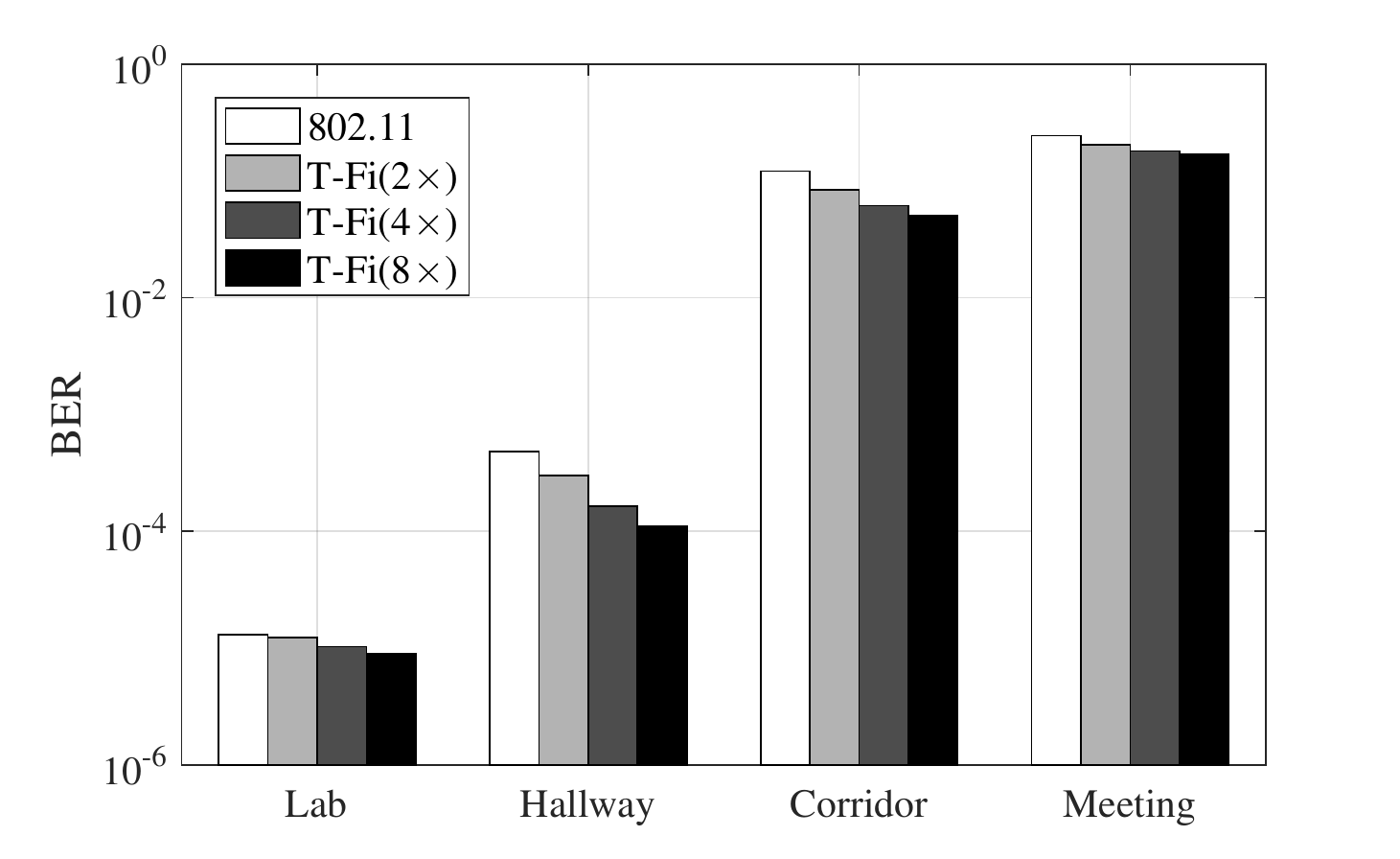}}\hspace{-0.2cm}
		\subfigure[\scriptsize Different modulations]
		{\label{fig:loc_mod}\includegraphics[width=0.5\textwidth]{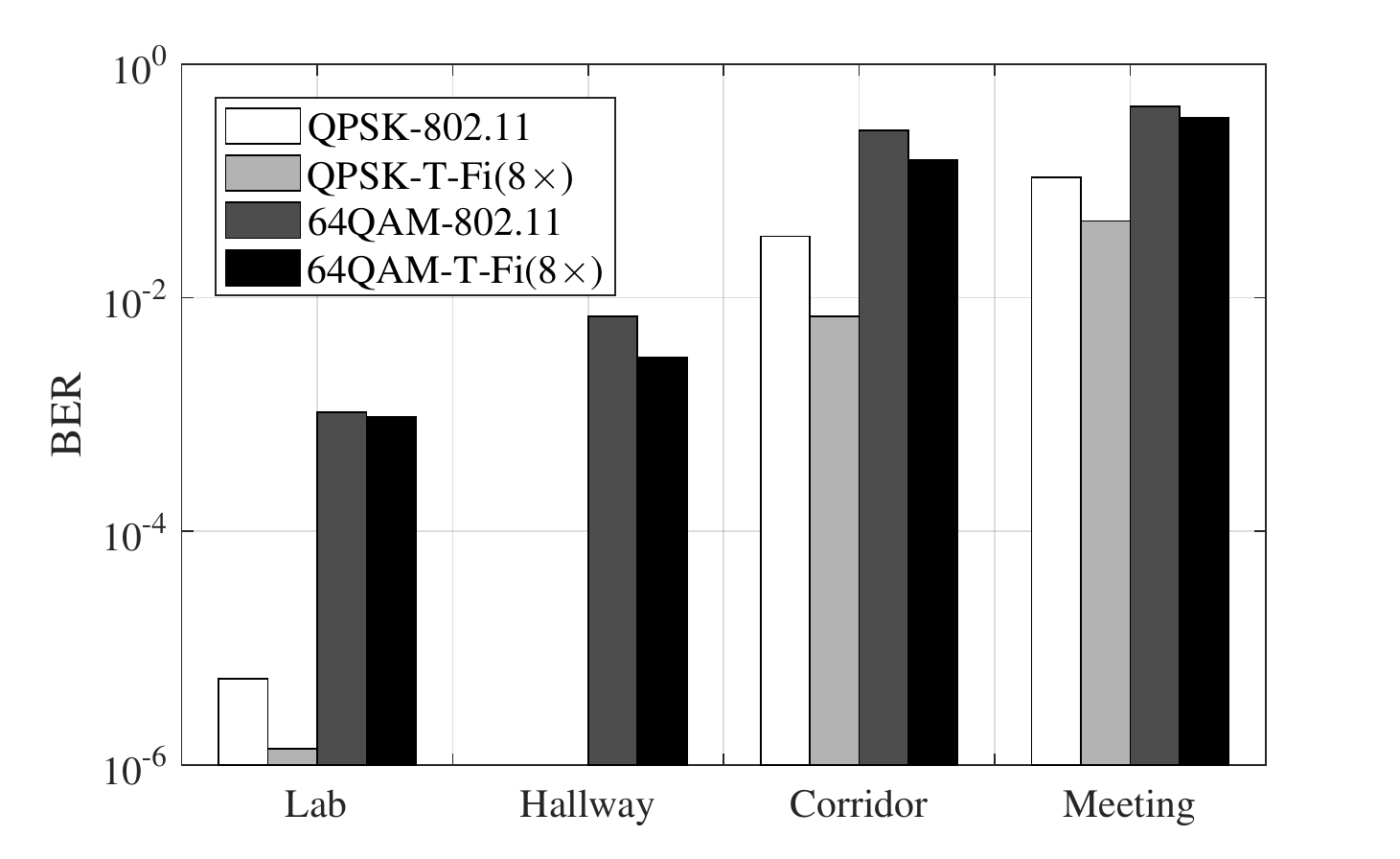}}\vspace{0.3cm}
		\caption{BER in different locations.}
		\label{fig:loc}
	\end{minipage} \vspace{0.2cm}
	\hspace{-0.2cm}
	\begin{minipage}[b]{0.33\textwidth}\centering
		\includegraphics[width=1\textwidth]{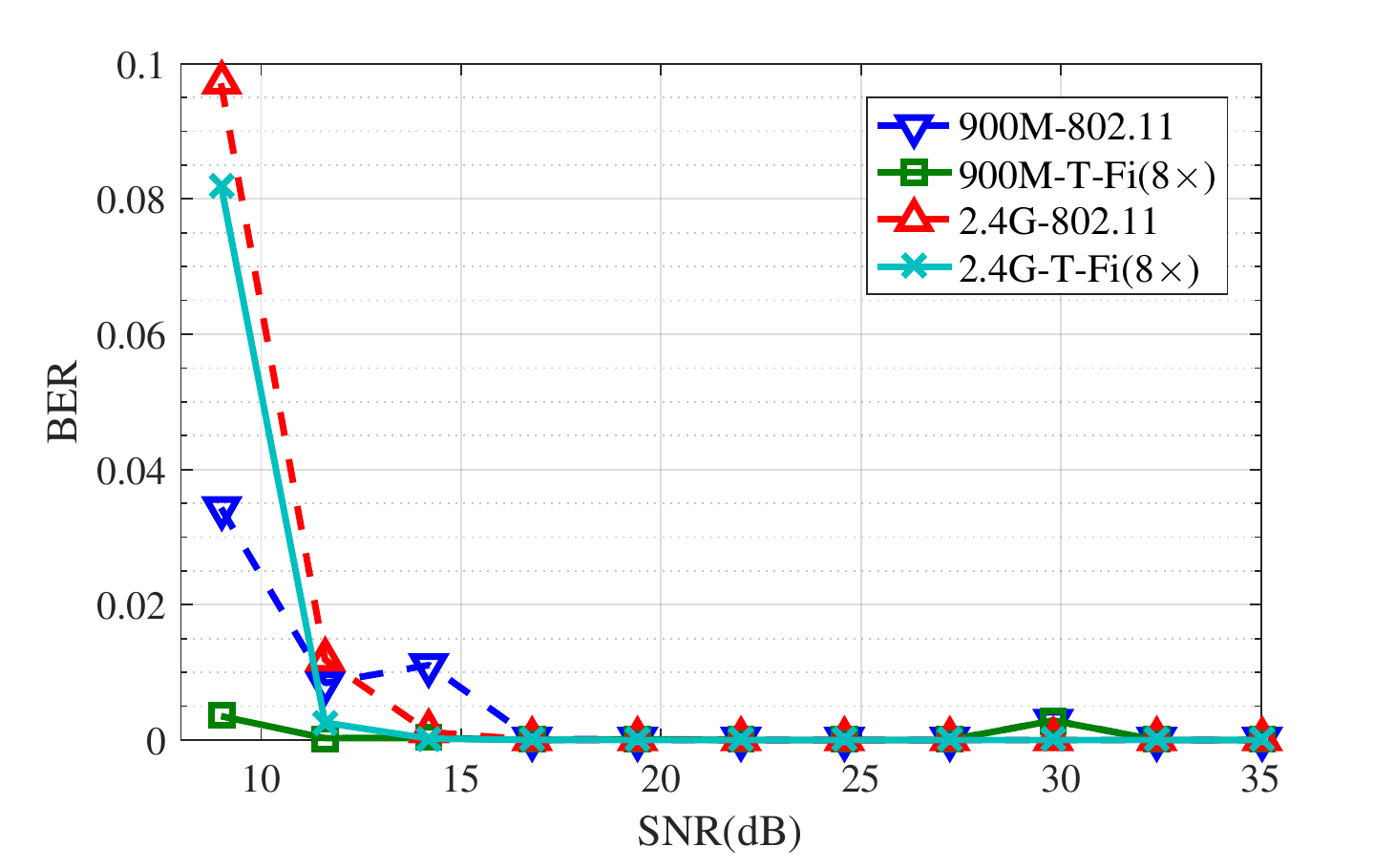}\vspace{0.45cm}
		\caption{BER under different carrier frequencies.} \label{fig:freq}
	\end{minipage} \vspace{0.2cm}
\end{figure*}

Figure~\ref{fig:loc_rate} compares the decoding performance under different clock rates at different locations. For all locations, \texttt{T-Fi} outperforms the standard 802.11 decoding, and higher clock rate \texttt{T-Fi} receivers achieve lower BER than the lower clock rate ones. Compared to the standard receiver, \texttt{T-Fi} at $8\times$ clock rate reduces BER to 56\%-23\% across all locations. \texttt{T-Fi} yields higher performance gain in the hallway and the corridor. The reason behind this is that the BER in these two locations falls in a sweet range where there is room for BER improvement while noise can be amortized by exploiting overclocking. The results confirm that \texttt{T-Fi} achieves stable performance gain over a wide range of wireless environment. Figure~\ref{fig:loc_mod} further shows the BER performance for different modulation schemes at the four locations. Though the BER of different modulations varies significantly across different locations, \texttt{T-Fi} achieves steady BER gain over the standard receiver.

\textbf{Impact of carrier frequency.} Previous experiments use 900~MHz as the carrier frequency, which is the band specified by IEEE 802.11ah. Recently, several leading companies have been exploring the possibility of using 2.4~GHz for IoT transmissions. To meet this potential demands, we now evaluate the BER performance in both carrier frequencies under various SNR conditions.

Figure~\ref{fig:freq} shows the BER for 16QAM decoding. We have the following observations: (i) The BER performance at both frequencies has the same trend and is comparable in most cases; (ii) \texttt{T-Fi} achieves lower BER compared to the standard receiver at both frequencies. The results validate the feasibility of \texttt{T-Fi} at both carrier frequencies.

\vspace{0.3cm}
\section{Related Work}\label{sec:relatedwork}
\textbf{IoT Wi-Fi standard.} IEEE 802.11ah, known as Wi-Fi HaLow~\cite{wifihalow}, is the first and only Wi-Fi protocol dedicatedly designed for IoT. The target of this IoT Wi-Fi protocol is to utilize the Wi-Fi technology to provide low power and long range transmission for the emerging low-end IoT devices. It reuses the OFDM PHY as in IEEE 802.11ac, but provides a bundle of low power features, particularly at the MAC layer. Research efforts have focused on enhancing the MAC layer~\cite{park2014enhancement,liu2013power}. Our goal is to design a transceiver architecture for IoT Wi-Fi networks that (i) completely conforms to the IoT Wi-Fi standard without any protocol modifications, and at the same time (ii) take full advantage of AP to further reduce the TX power of IoT devices. Our compliant design can be seamlessly integrated into the existing IoT Wi-Fi protocol. Existing Wi-Fi power models~\cite{qiao2003miser,jung2002energy,baiamonte2006saving,bruno2002optimization,carvalho2004modeling,chen2008edca,chen2008edca,ergen2007decomposition,garcia2011energy} show that power amplifier operations dominate the total power consumption of Wi-Fi radios. Thus, the saved power budget by our design can facilitate lower power or longer range transmission, which fully aligns with the targets of IoT Wi-Fi protocols.

\textbf{Downclocking.} Recently, researchers have demonstrated that downclocking Wi-Fi radios can effectively reduce the energy consumption during packet reception and idle listening. E-Mili~\cite{mobicom11emili} pioneers this kind of mechanisms to downclock receiver's clock rate during idle listening, and switches to full clock rate for packet reception. SloMo~\cite{nsdi13slomo} enables packet decoding in IEEE 802.11b while downclocking by exploiting the sparsity in direct sequence spread spectrum (DSSS) PHY. SEER~\cite{wang2017wideband} designs a special preamble to allow narrowband receiver sensing a much wider band signal without boosting the sampling rate. Downclocking for packet reception has been extended to OFDM-based Wi-Fi by leveraging the gap between modulation and SNR~\cite{mobicom14enfold}. Instead of exploiting PHY redundancy, Sampleless Wi-Fi~\cite{wang2017sampleless,wang2016rateless} utilizes the retransmission opportunities to decode packets at downclocked rates without relying on PHY redundancy, and thus makes downclocking OFDM packet reception feasible under low SNR conditions. \texttt{T-Fi} takes an opposite approach by overclocking the receiver's radio to reduce the power consumption of the transmitter, which is complementary to downclocking approaches by bringing low power to the transmitter side.

\vspace{0.3cm}
\section{Conclusion}\label{sec:conclusion}
This paper introduces \texttt{T-Fi}, an asymmetric transceiver paradigm for IoT that pushes power burden to the AP side, and enables IoT devices to transmit packets at power levels that are even lower than the minimal power required by conventional receivers. We think this is an important design point in IoT communications, and the teeter-totter fashion in \texttt{T-Fi} has significant ramifications in the new transceiver design for IoT Wi-Fi protocols. Our experimental evaluation confirms the benefits of \texttt{T-Fi} in real environments. We hope the design can contribute the wireless community by providing some insights for future transceiver design that takes advantage of the hardware asymmetry between APs and IoT devices.

\vspace{0.3cm}
\section*{Acknowledgment}

The research was supported in part by the National Science Foundation of China under Grant 61502114, 91738202, 61729101, and 61531011, Major Program of National Natural Science Foundation of Hubei in China with Grant 2016CFA009, and the Fundamental Research Funds for the Central Universities with Grant number 2015ZDTD012.

\balance
\bibliographystyle{IEEEtran}
\bibliography{IEEEabrv,./oversample}

\end{document}